\documentclass[letter]{aa} 
\usepackage{graphicx}
\usepackage{txfonts}
 \usepackage{longtable}
\begin{document}
\title{Ortho-to-para ratio of interstellar heavy water\thanks{Herschel
    is an ESA space observatory with science instruments provided by
    European-led principal Investigator consortia and with important
    participation from NASA}}


   \author{
Vastel C. \inst{\ref{cesr},\ref{cesr2}} \and
Ceccarelli C. \inst{\ref{laog},\ref{bordeaux},\ref{bordeaux2}} \and
Caux E. \inst{\ref{cesr},\ref{cesr2}} \and
Coutens A. \inst{\ref{cesr},\ref{cesr2}} \and
Cernicharo J. \inst{\ref{cab}} \and
Bottinelli S. \inst{\ref{cesr},\ref{cesr2}} \and
Demyk K. \inst{\ref{cesr},\ref{cesr2}} \and
Faure A. \inst{\ref{laog}} \and
Wiesenfeld L. \inst{\ref{laog}} \and
Scribano Y. \inst{\ref{licb}} \and
Bacmann A.\inst{\ref{laog},\ref{bordeaux},\ref{bordeaux2}} \and
Hily-Blant P. \inst{\ref{laog}} \and
Maret S. \inst{\ref{laog}} \and
Walters A. \inst{\ref{cesr},\ref{cesr2}} \and
Bergin E.A. \inst{\ref{annarbor}} \and
Blake G.A. \inst{\ref{caltech}} \and
Castets A. \inst{\ref{laog},\ref{bordeaux},\ref{bordeaux2}} \and
Crimier N. \inst{\ref{laog},\ref{cab}} \and
Dominik C. \inst{\ref{uva},\ref{nijmegen}} \and
Encrenaz P. \inst{\ref{lerma}} \and
Gerin M. \inst{\ref{lerma}} \and
Hennebelle P. \inst{\ref{lerma}} \and
Kahane C. \inst{\ref{laog}} \and
Klotz A. \inst{\ref{cesr},\ref{cesr2}} \and
Melnick G. \inst{\ref{cfa}} \and
Pagani L. \inst{\ref{lerma}} \and
Parise B. \inst{\ref{bonn}} \and
Schilke P. \inst{\ref{bonn},\ref{koln}} \and
Wakelam V. \inst{\ref{bordeaux},\ref{bordeaux2}} \and
Baudry A. \inst{\ref{bordeaux},\ref{bordeaux2}} \and
Bell T. \inst{\ref{caltech}} \and
Benedettini M. \inst{\ref{ifsi}} \and
Boogert A. \inst{\ref{ipac}} \and
Cabrit S. \inst{\ref{lerma}} \and
Caselli P. \inst{\ref{leeds}} \and
Codella C. \inst{\ref{arcetri}} \and
Comito C. \inst{\ref{bonn}} \and
Falgarone E. \inst{\ref{lerma}} \and
Fuente A. \inst{\ref{oan}} \and
Goldsmith P.F. \inst{\ref{jpl}} \and
Helmich F. \inst{\ref{sron}} \and
Henning T. \inst{\ref{heidelberg}} \and
Herbst E. \inst{\ref{ohio}} \and
Jacq T. \inst{\ref{bordeaux},\ref{bordeaux2}} \and
Kama M. \inst{\ref{uva}} \and
Langer W. \inst{\ref{jpl}} \and
Lefloch B. \inst{\ref{laog}} \and
Lis D. \inst{\ref{caltech}} \and
Lord S. \inst{\ref{ipac}} \and
Lorenzani A. \inst{\ref{arcetri}} \and
Neufeld D. \inst{\ref{hopkins}} \and
Nisini B. \inst{\ref{oar}} \and
Pacheco S. \inst{\ref{laog}} \and
Pearson J. \inst{\ref{jpl}} \and
Phillips T. \inst{\ref{caltech}} \and
Salez M. \inst{\ref{lerma}} \and
Saraceno P. \inst{\ref{ifsi}} \and
Schuster K. \inst{\ref{iram}} \and
Tielens X. \inst{\ref{leiden}} \and
van der Tak F. \inst{\ref{sron},\ref{kaypten}} \and
van der Wiel M.H.D. \inst{\ref{sron},\ref{kaypten}} \and
Viti S. \inst{\ref{ucl}} \and
Wyrowski F. \inst{\ref{bonn}} \and
Yorke H. \inst{\ref{jpl}}
Cais, P.  \inst{\ref{bordeaux},\ref{bordeaux2}} \and
Krieg, J.M. \inst{\ref{lerma}} \and
Olberg, M. \inst{\ref{sron},\ref{chalm}} \and
Ravera, L. \inst{\ref{cesr},\ref{cesr2}} 
}

   \institute{
Centre d'Etude Spatiale des Rayonnements, Universit\'e Paul Sabatier, Toulouse, France
\label{cesr}
\and CNRS/INSU, UMR 5187, Toulouse, France
\label{cesr2}
\and Laboratoire d'Astrophysique de Grenoble, UMR 5571-CNRS, Universit\'e Joseph Fourier, Grenoble, France
\label{laog}
\and Universit\'{e} de Bordeaux, Laboratoire d'Astrophysique de Bordeaux, Floirac, France
\label{bordeaux}
\and CNRS/INSU, UMR 5804, Floirac cedex, France
\label{bordeaux2}
\and Centro de Astrobiolog\`{\i}a, CSIC-INTA, Madrid, Spain
\label{cab}
\and Laboratoire Interdisciplinaire Carnot de Bourgogne, UMR 5209-CNRS, Dijon Cedex, France\\
\label{licb}
\and Department of Astronomy, University of Michigan, Ann Arbor, USA
\label{annarbor}
\and California Institute of Technology, Pasadena, USA
\label{caltech}
\and Astronomical Institute 'Anton Pannekoek', University of Amsterdam, Amsterdam, The Netherlands
\label{uva}
\and Department of Astrophysics/IMAPP, Radboud University Nijmegen,  Nijmegen, The Netherlands
\label{nijmegen}
\and Laboratoire d'Etudes du Rayonnement et de la Mati\`ere en Astrophysique, UMR 8112  CNRS/INSU, OP, ENS, UPMC, UCP, Paris, France
\label{lerma}
\and Harvard-Smithsonian Center for Astrophysics, Cambridge MA, USA
\label{cfa}
\and Max-Planck-Institut f\"{u}r Radioastronomie, Bonn, Germany
\label{bonn}
\and Physikalisches Institut, Universit\"{a}t zu K\"{o}ln, K\"{o}ln, Germany
\label{koln}
\and INAF - Istituto di Fisica dello Spazio Interplanetario, Roma, Italy
\label{ifsi}
\and Infared Processing and Analysis Center,  Caltech, Pasadena, USA
\label{ipac}
\and School of Physics and Astronomy, University of Leeds, Leeds UK
\label{leeds}
\and INAF Osservatorio Astrofisico di Arcetri, Florence Italy
\label{arcetri}
\and IGN Observatorio Astron\'{o}mico Nacional, Alcal\'{a} de Henares, Spain
\label{oan}
\and Jet Propulsion Laboratory,  Caltech, Pasadena, CA 91109, USA
\label{jpl}
\and SRON Netherlands Institute for Space Research, Groningen, The Netherlands
\label{sron}
\and Max-Planck-Institut f\"ur Astronomie, Heidelberg, Germany
\label{heidelberg}
\and Ohio State University, Columbus, OH, USA
\label{ohio}
\and Johns Hopkins University, Baltimore MD,  USA
\label{hopkins}
\and INAF - Osservatorio Astronomico di Roma, Monte Porzio Catone, Italy
\label{oar}
\and Institut de RadioAstronomie Millim\'etrique, Grenoble - France
\label{iram}
\and Leiden Observatory, Leiden University, Leiden, The Netherlands
\label{leiden}
\and Kapteyn Astronomical Institute, University of Groningen, The Netherlands
\label{kaypten}
\and Department of Physics and Astronomy, University College London, London, UK
\label{ucl}
\and Chalmers University of Technology, G{\o}terborg, Sweden
\label{chalm}
}

   \date{Received ; accepted }

 
  \abstract
  {Despite the low elemental deuterium abundance in the Galaxy,
    enhanced molecular D/H ratios have been found in the environments
    of low-mass star forming regions, and in particular the Class 0
    protostar IRAS 16293-2422.}
  {The CHESS (Chemical HErschel Surveys of Star forming regions) Key
    Program aims at studying the molecular complexity of the
    interstellar medium. The high sensitivity and spectral resolution
    of the HIFI instrument provide a unique opportunity to observe the
    fundamental 1$_{1,1}$ -- 0$_{0,0}$ transition of the ortho--D$_2$O
    molecule, inaccessible from the ground, and to determine the
    ortho-to-para D$_2$O ratio.
 }
 {We have detected the fundamental transition of the ortho-D$_2$O
   molecule at 607.35 GHz towards IRAS 16293-2422. The line is seen in
   absorption with a line opacity of 0.62 $\pm$ 0.11 ($1
     \sigma$). From the previous ground-based observations of the
   fundamental 1$_{1,0}$ --1$_{0,1}$ transition of para--D$_2$O seen
   in absorption at 316.80 GHz we estimate a line opacity of 0.26 $\pm$ 0.05
     ($1 \sigma$).}
 { We show that the observed absorption is caused by the cold gas in
   the envelope of the protostar.  Using these new observations, we
   estimate for the first time the ortho to para D$_2$O ratio to
     be lower than 2.6 at a 3 $\sigma$ level of uncertainty, to be
   compared with the thermal equilibrium value of 2:1.  }
   {}


   \maketitle
%

\section{Introduction}
\begin{figure}[b]
  \centering
  \includegraphics[width=7cm]{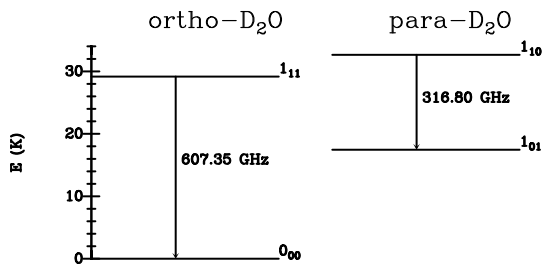}
  \caption{Energy levels for the detected fundamental lines of D$_2$O.}
  \label{transitions}
\end{figure}
Among all molecules in interstellar space, water is a special one
because of its dominant role in the cooling of warm gas and in the
oxygen chemistry as well as for its role in the chemistry of the
atmospheres of exoplanets and its potential connection with life.
Water abundance in cold molecular gas is very low because it is
frozen onto the interstellar grains and forms icy mantles around
them. 
Although water can form theoretically via gaseous reactions which first form H$_2$O$^+$ and H$_3$O$^+$ (e.g. Rodgers \& Charnley 2002), 
no observational evidence has been collected so far. It is believed that the major mechanism of water formation is on 
grain surfaces.
One observable that helps to discriminate between the various
formation mechanisms is the abundance of single and double deuterated
water with respect to the normal isotopologue.  Another potential
discriminant can be the ortho-to-para ratio (OPR), namely the ratio
between water molecules with different nuclear spins.  In fact, since
radiative and inelastic collisional transitions between the two ortho
and para states are strongly forbidden, the OPR is set at the moment
of the water formation and it is changed by nuclear spin reactions
exchange later on. This can occur either in the gas phase by reactions
with ions in which actual nuclei change places, or on the grain
surfaces by interaction with electron spins or, perhaps, even other
nuclear spins (e.g. Le Bourlot et al. 2000, Limbach et al. 2006).
Although little is known on the spin exchange in the gas phase, it is
usually assumed that this is a slow process and that the OPR is likely
to keep memory of the moment of its formation.
Emprechtinger et al. and Lis et al. in this volume report
determinations of the water OPR in several environments based on new
Herschel observations.  The doubly deuterated isotopologue of water,
D$_2$O, consists of two species, ortho and para with a nuclear spin
statistic weight 2:1. So far D$_2$O has only been detected towards the
solar type protostar IRAS 16293-2422 (hereafter IRAS16293), via the
observation of the fundamental transition of the para--D$_2$O
transition at 316.8 GHz (see our Figure \ref{transitions}; Butner et
al. 2007). The observed line profile (see Figure \ref{figures}) shows
a component in emission with a deep absorption at the cloud velocity
($\sim$ 4 km~s$^{-1}$). The emission component has been attributed to
heavy water in the hot corino of this source where the grain ices are
sublimated and released into the gas phase (Ceccarelli et al. 2000;
Bottinelli et al. 2004), based on the detailed analysis of several HDO
lines observed in IRAS16293 (Parise et al. 2005).  The absorption
component, whose linewidth is 0.5 km~s$^{-1}$, is likely due to the
foreground gas (molecular cloud and cold envelope). Therefore, the
absorption component provides a straightforward measure of the column
density of para--D$_2$O in the cold gas surrounding IRAS16293.


\section{Observations and results}\label{sec:observations-results}

\begin{figure}
  \centering
  \includegraphics[width=8cm]{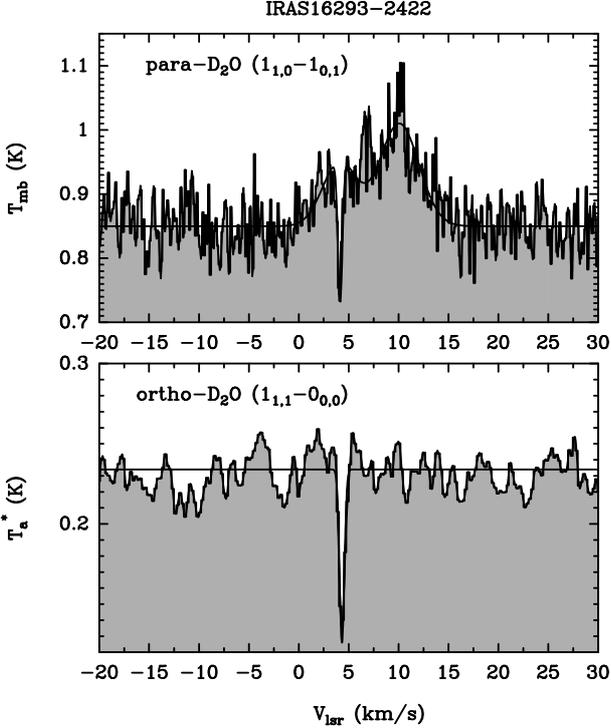}
  \caption{Profile of the para--D$_2$O (1$_{1,0}$-1$_{0,1}$) line
    (histogram) observed at JCMT (upper panel), as well as the 3
    components gaussian fit (solid line) and ortho--D$_2$O
    (1$_{1,1}$-1$_{0,1}$) line observed with HIFI (bottom panel).}
  \label{figures}
\end{figure}

In the framework of the Key Program CHESS (Ceccarelli et al., this
volume), we observed the solar type protostar IRAS16293 with the HIFI
instrument (de Graauw et al., 2010; Roelfsema et al., this volume) on
board the Herschel Space Observatory (Pilbratt et al., 2010).  A full
spectral coverage of band 1b between 554.5 and 636.5\,GHz was
performed on March 2nd 2010, using the HIFI Spectral Scan Double Beam
Switch (DBS) mode with optimization of the continuum.  The fundamental
ortho--D$_2$O (1$_{1,1}$-0$_{0,0}$) transition lies in this frequency
range, at 607.35 GHz (see Figure \ref{transitions}).  The HIFI Wide
Band Spectrometer (WBS) was used, providing a spectral resolution of
1.1 MHz ($\sim$0.55 km s$^{-1}$ at 600 GHz) over an instantaneous
bandwidth of 4~x~1\,GHz. Note that the data are acquired at the
  Nyquist sampling, therefore, with 0.5 MHz steps. The targeted
coordinates were $\alpha_{2000}$ = 16$^h$ 32$^m$ 22$\fs$75,
$\delta_{2000}$ = $-$ 24$\degr$ 28$\arcmin$ 34.2$\arcsec$. The beam
size at 610~GHz is about 35$\arcsec$, the theoretical main beam
(resp. forward) efficiency is 0.72 (resp. 0.96), and the DBS reference
positions were situated approximately 3$\arcmin$ east and west of the
source.  The data have been processed using the standard HIFI pipeline
up to level 2 with the ESA-supported package HIPE 3.01 (Ott et
al. 2010).  The 1 GHz chunks are then exported as fits files into
CLASS/GILDAS format\footnote{http://www.iram.fr/IRAMFR/GILDAS} for
subsequent data reduction and analysis using generic spectral survey
tools developed in CLASS in our group. When present, spurs were
removed in each 1 GHz scan and a low order polynomial ($\leq 2$)
  baseline was fitted over line-free regions to correct residual
bandpass effects. These polynomials were subtracted and used to
determine an accurate continuum level by calculating their
medians. Sideband deconvolution is computed with the minimisation
algorithm of Comito \& Schilke (2002) implemented into CLASS using the
baseline subtracted spectra and assuming side-band gain ratio to be
unity for all tunings. Both polarisations were averaged to lower the
noise in the final spectrum. The continuum values obtained are well
fitted by straight lines over the frequency range of the whole
band. The single side band continuum derived from the polynomial fit
at the considered frequency was added to the spectra.  Finally, the
deconvolved data were analysed with CASSIS software\footnote{Developed by
  CESR-UPS/CNRS: http://cassis.cesr.fr}.  Exact measurements of the main beam efficiency have
not been performed on planets at the time of our
observations. However, we are dealing with absorption measurements,
and are only interested in the relative depth of the absorption
compared to the continuum level. Consequently we present in the
following the spectrum (Figure 2) and parameters (Table 1) in T$_a^*$
for the ortho--D$_2$O line.  The bottom panel of Figure \ref{figures}
shows the resulting HIFI spectrum, with the measured continuum level
of (234 $\pm$ 19) mK (where the error includes the statistical
  error only).  Note that the absolute calibration doesn't matter for
  absorption, since lines and continuum are affected the same
  way. Therefore, the main source of the uncertainty is the accuracy
  in the continuum.  The achieved rms is about 12 mK in T$_a^*$, in
the 0.5 MHz frequency bin. The fundamental ortho--D$_2$O transition at
607349.449 MHz is well detected in absorption against the strong
continuum, at the velocity of $\sim$ 4 km/s.  No other lines in
  the image sideband are expected at this velocity. The parameters of
the line, obtained using CASSIS, which takes into account the ortho
and para D$_2$O forms separately from the Cologne Database for
Molecular Spectroscopy (M\"uller et al. 2005, Br\"unken et al. 2007),
are reported in Table 1. We report in the same table also the
parameters of the para--D$_2$O (1$_{1,0}$-1$_{0,1}$) fundamental line
previously observed at the JCMT, published by Butner et al. (2007), at
a rest frequency of 316799.81 MHz.  The data were retrieved from the
JCMT archive and reduced within CLASS. We performed a 3-component
gaussian fit with CASSIS and the resulting fit is reproduced in Figure
\ref{figures} on top of the data in main beam temperatures. The
para--D$_2$O line in emission has an intensity of 0.10 $\pm$ 0.02 K in
main beam temperature, and a linewidth of 4.01 $\pm$ 0.77
km~s$^{-1}$. The bright line at a V$_{lsr}$ of 10.1 km~s$^{-1}$ is likely due to
CH$_3$OD (see Butner et al. 2007) with an intensity of $\rm (0.16 \pm 0.01) K$ 
and a linewidth of $\rm (4.6 \pm 0.5) km~s^{-1}$. The parameters for the resulting fit of the para--D$_2$O 
absorption line are quoted in Table 1. 
\begin{table*}[tb]
  \caption{Derived parameters of the ortho and para D$_2$O fundamental
    lines. Note that the parameters are in T$_a^*$ for ortho--D$_2$O 
    and T$_{mb}$ for para--D$_2$O (see text).}    
\label{table1}      
\centering                          
\begin{tabular}{c c c c c c c c c c}        
\hline\hline                 
Species      &  Transition & Frequency & Telescope & $\int T dv$ & T$_{abs}$ = T$_C$-T$_L$  & $\Delta$V & V$_{LSR}$  & T$_C$ & $\tau$ \\    
	           &                     & GHz            &                      & (mK km/s) & (mK) & (km/s) & (km/s) & (mK) & \\
\hline                        
ortho--D$_2$O & 1$_{1,1}$--0$_{0,0}$ &   607.34945  & Herschel &  77 $\pm$ 17    & 108 $\pm$ 11       & 0.57 $\pm$ 0.09  & 4.33 $\pm$ 0.04  & 234 $\pm$ 19 & 0.62 $\pm$ 0.11\\
para--D$_2$O  & 1$_{1,0}$--1$_{0,1}$ &   316.79981  & JCMT  & 120 $\pm$ 49 & 220  $\pm$ 30       & 0.55 $\pm$ 0.15  & 4.15 $\pm$ 0.04  & 850 $\pm$ 35 & 0.26 $\pm$ 0.05\\
 \hline                                   
\end{tabular}
\end{table*}

\section{Determination of the D$_2$O OPR}\label{sec:determ-d_2o-opr}

Crimier et al. (2010) have used the JCMT SCUBA maps of IRAS16293 at
450 $\mu$m and 850 $\mu$m (and other data) to reconstruct the
structure of the IRAS16293 envelope.  From this work, one can compute
the expected continuum in the HIFI beam at 607 GHz (o--D$_2$O
line). Using the SED of Crimier et al. (Fig. 1 panel d) and their
Table 1, the IRAS16293 flux is $270 \pm 108$ Jy at 450 $\mu$m and the
HIFI beam contains approximately 80\% of the total source flux (Fig 1,
panel b). One can note that the SED steep slope yields the flux at 607
GHz to be smaller than the one at 450 $\mu$m ($\sim\,660$ GHz) by
about 30\%, making the expected flux at 607 GHz to be about $0.7
\times 0.8 \times (270 \pm 108)$ Jy i.e. ($0.34\pm 0.14$) K, using the
HIFI Jy to K conversion factor (C. Kramer : Spatial response,
contribution to the HIFI framework document), in perfect agreement
with the observed continuum value ($\sim$ 0.33 K in main beam
temperature).  Most of the continuum, more than 70\% (resp. 80\%) of
its peak emission at 316 GHz (resp. 607 GHz) is emitted from a region
of about 900 AU in radius ($\sim$ 15$^{\arcsec}$ in diameter). The
absorption of the continuum by heavy water is most likely due to the
cold envelope surrounding IRAS16293 as well as the parent cloud, much
more extended than the continuum emitting region. Note that, as
  far as the sizes of the absorbing layer are larger than the sizes of
  the region emitting the continuum, the line-to-continuum ratio does
  not depend on the sizes of the telescope beam used for the
  observations. Therefore, we can compute the D$_2$O OPR directly from
  the line-to-continuum ratios of the JCMT and Herschel observations,
  with no further correction.  Note also that the para--D$_2$O line
has an emission component that Butner et al. (2007) attributed to the
hot corino region, whereas here we are dealing with an {\it
  absorption} component only. On the contrary, the ortho--D$_2$O line
reported here shows an absorption only as the emission component is
very likely diluted in the 35$^{\arcsec}$ HIFI beam, much larger than
the 15$^{\arcsec}$ JCMT beam at 316 GHz.  

\begin{figure}[b]
  \centering
  \includegraphics[bb=20 20 480 310,width=8cm]{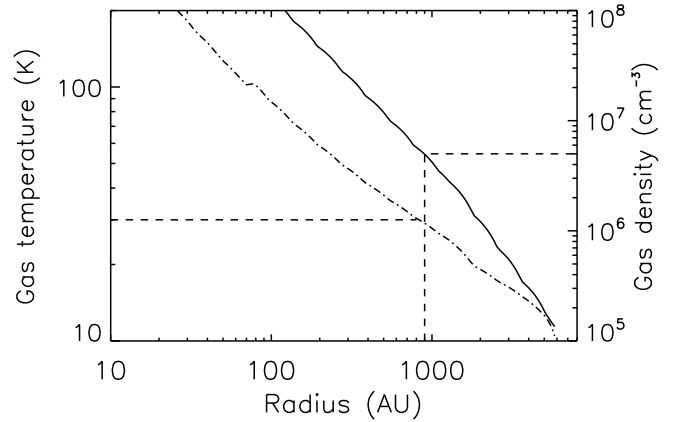}
  \caption{Density profile (solid line) and gas temperature profile
    (dot-dashed) of the IRAS16293 envelope, as computed by Crimier et
    al. (2010). Values at a radial distance of about 900 AU are also
    indicated (see text). A distance of 120 pc has been used in this
    recent computation (Loinard et al. 2008).}
  \label{profile}
\end{figure}

Adopting the density and temperature profiles of the envelope of
IRAS16293 (Crimier et al. 2010), the gas at a distance larger (in
radius) than 900 AU has a temperature lower than 30 K and a density
lower than about $5\times10^6$ cm$^{-3}$ (see Figure \ref{profile}).
Thus, given the temperature of the gas absorbing the D$_2$O lines, we
only consider the first two levels of each D$_2$O form.  We use the
recently computed collisional rates for the two fundamental
deexcitation transitions of ortho and para--D$_2$O with para--H$_2$ in
the 10-30 K range: $2.3\times10^{-11}$ and $3.8\times10^{-11}$
cm$^3$~s$^{-1}$ respectively (Wiesenfeld, Faure \& Scribano, in
preparation).  At the low temperatures found in the cold envelope, it
is likely that H$_2$ is mainly in its para form (Pagani et al. 2009,
Troscompt et al. 2009).  With the collisional rates given above, the
critical densities of the ortho- and para- D$_2$O fundamental
transitions are $1\times10^8$ and $2\times10^7$ cm$^{-3}$
respectively, and the upper levels of the two transitions are only
moderately sub-thermally populated for a density of 5 $\times$ 10$^6$
cm$^{-3}$. For a two-level system, the species column density can be
computed as follows:
\begin{equation}
\rm N_{tot}=\frac{8\pi\nu^{3}}{A_{ul}c^{3}}\Delta V \frac{\sqrt{\pi}}{2\sqrt{ln2}}\tau\frac{Q(T_{ex})}{g_{u}}\frac{\exp(E_u/kT_{ex})}{[\exp(h\nu/kT_{ex})-1]}
\end{equation}
where A$_{ul}$ is the Einstein coefficient ($2.96\times10^{-3}$
s$^{-1}$ for the ortho transition and $6.3\times10^{-4}$ s$^{-1}$ for
the para transition), E$_{u}$ is the upper level energy (E$_u$/k=15.2
K for the para transition and =29.2 K for the ortho transition),
g$_{u}$ is the upper statistical weight (3 $\times$ (2J+1) for the
para transition, 6 $\times$ (2J+1) for the ortho transitions), $\nu$
is the frequency (316.79981 GHz for the para transition and 607.349449
GHz for the ortho transition), $\Delta V$ is the linewidth (cm
s$^{-1}$) and $\tau$ is the opacity at the line center.  T$_{ex}$ is
the excitation temperature and Q(T$_{ex}$) is the corresponding
partition function.  In the approximation of the escape probability
formalism, T$_{ex}$ is defined by the equation:
\begin{equation}
\rm T_{ex}~=~\frac{h\nu/k}{h\nu/kT_k+\ln(1+A_{ul}\beta/C_{ul})}
\end{equation}
where $\rm C_{ul}~=~\gamma_{ul} \times n_{collision}$, $\rm
n_{collision}$ being the density of the collision partner (in this
case para--H$_2$) and $\rm \gamma_{ul}$ being the collisional rate in
cm$^{3}$~s$^{-1}$ (values given above). The $\beta$ parameter
represents the probability that a photon at some position in the cloud
escapes the system. For a static, spherically symmetric and
homogeneous medium, Osterbrock and Ferland (2006, Appendix 2) derives
this parameter as a function of the optical depth $\rm \tau$ in the
direction of the observer. The opacity at the line center is expressed
as a function of the line depth (T$_{abs}$ = T$_C$-T$_L$) and the
continuum (T$_{C}$):
\begin{equation}
\rm \tau=-\ln\left(1-\frac{T_{abs}}{T_{C}-J_\nu(T_{ex})+J_{\nu}(T_{cmb}))}\right)
\end{equation}
%
Where $\rm J_\nu(T_{ex})=(h\nu/k)/(exp(h\nu/k)-1) $ and
T$_{cmb}$ is the cosmic microwave background radiation temperature
(2.73 K).  In the limit of $\tau \gg 1$, $\rm T_{C} - T_{abs} \sim
J_\nu(T_{ex})-J_{\nu}(T_{cmb})$, and $\rm T_{ex} \sim 5 K$. Since the
D$_2$O transitions are probably optically thin, we can reasonably
assume that $\rm T_{ex}$ is lower than 5 K and
$\rm J_\nu(T_{ex})-J_{\nu}(T_{cmb})$ is negligible.

\begin{figure}
  \centering
  \includegraphics[bb=70 10 494 340,width=7cm]{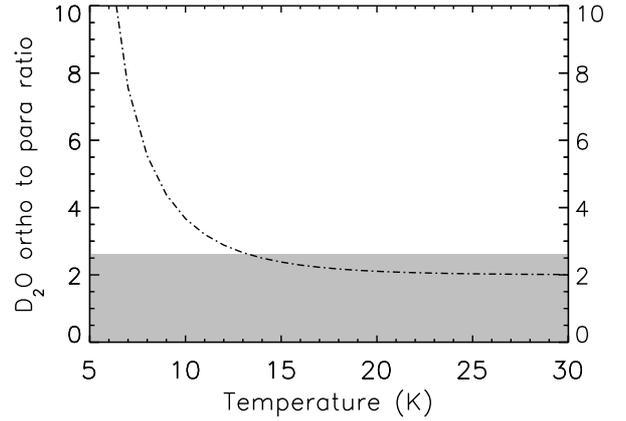}
  \caption{Upper limit on the measured D$_2$O OPR (2.6, see text) as a grey box and
    the Boltzmann value (dotted-dashed line) as a function of  temperature.}
  \label{opratio}
\end{figure}

As discussed above, we assume that the absorbing layer is much larger
than the continuum emitting region. Considering the uncertainty on the
H$_2$ density (lower than 5 $\times$ 10$^6$ cm$^{-3}$) and the kinetic
temperature (lower than 30 K), we applied the method described above
to determine the column densities with $\rm n_{H_2} = 10^6 cm^{-3}$
and $\rm T_{kin} \sim 20 K$.  Table \ref{table1} lists the
  computation of the optical depths for both lines as well as the
  corresponding uncertainties. Since $\rm \tau = -ln(T_L/T_C)$ the
  uncertainty in the line optical depth is
  given by $\rm \delta \tau = exp(\tau) \times \delta (T_L/T_C)$.
Our computation yields an OPR equal to 1.1 $\pm$ 0.4 with the
corresponding column densities: N$_{ortho}$ = (8.7 $\pm$ 2.1)
10$^{11}$ cm$^{-2}$ and N$_{para}$ = (7.8 $\pm$ 2.6) 10$^{11}$
cm$^{-2}$.  All errors here are 1 $\sigma$). Both lines are
optically thin and their T$_{ex}$ are lower than 5 K.  Note that
decreasing the density and/or the kinetic temperature doesn't change
the OPR by more than 10\%. Therefore, the OPR is lower than 2.4
  at a 3 $\sigma$ level of uncertainty (where we added the 3 $\sigma$
  statistical error and the mentioned 10\% to the 1.1 value).  
We assumed (see section 2) that the relative gains on the lower and upper sidebands 
are equal. Since we do not have any information concerning the sideband ratio at the frequency of 
the D$_2$O line, we can only introduce a maximum uncertainty of $\rm 16\%$, corresponding to the 
overall calibration budget for band 1b. The resulting upper limit on the OPR is therefore increased to about 2.6.
Figure \ref{opratio} shows the measured OPR interval and the thermal
equilibrium as a function of the gas temperature.

\section{Conclusions} 
A discussed in \S \ref{sec:determ-d_2o-opr}, the gas absorbing
the D$_2$O line lies at more than 900 AU from the center and has a
temperature lower than 30 K.
The comparison between the upper value of the measured D$_2$O OPR and
the thermal equilibrium value shows that they are consistent with
  a gas at a temperature larger than about 15 K (at a 3 $\sigma$
  level of confidence), and, therefore, with the assumed absorbing gas
  location. On the other hand, the D$_2$O gas could have formed in a
previous phase, where the gas was colder, and, in this case, it
means that it had the time to thermalise to the Boltzmann value.
Unfortunately, given the poor knowledge of the mechanisms that
  can exchange the D$_2$O spins (see the Introduction), it is difficult
  to infer here the timescale for this change and, consequently, to
  give a lower limit to the object age. On the other hand, the relatively
  large uncertainty in the OPR derived here does not allow either to
  exclude a non- thermal equilibrium situation. Higher S/N
  observations will be needed to lower the uncertainty on the OPR
  value and give a more robust result.

Using the density and temperature profiles of the envelope of
IRAS16293 by Crimier et al. (2010), the column density of the gas
colder than 30 K is about $1\times10^{23}$ cm$^{-2}$. Therefore, the
D$_2$O abundance (with respect to H$_2$) is about $2\times10^{-11}$. 
An estimate of the water abundance profile will soon be available with the HIFI 
observations with a much higher spatial and spectral resolution than the one 
provided by the ISO observations (Ceccarelli et al. 2000). 
D$_2$O molecules might form with one OPR, but then could freeze out on grains
surfaces that could modify the ratio and then get desorbed.  Due to
the high uncertainty in the H$_2$O abundance, we cannot at the time
being completely exclude or confirm formation through grain surface
chemistry. A modeling of the OPR evolution is beyond the scope of the present letter. 
With an improved calibration and better understanding of the instrumental effects, 
a more accurate determination of the D$_2$O OPR in this source and potentially other 
sources will be possible. Also, ALMA may hopefully yield an answer in a near future with the 
observation of cold D$_2$O with a higher spatial resolution.

In summary, this Letter presents the first tentative to estimate the OPR for the
D$_2$O molecule, demonstrating the outstanding capabilities of the
HIFI instrument.  The poor knowledge of the mechanisms of
  exchange of the nuclear spins and the relatively large error in the
  derived OPR prevent to drawing firm conclusions on the formation of
  heavy water at that time.

\begin{acknowledgements}
  HIFI has been designed and built by a consortium of institutes and
  university departments from across Europe, Canada and the United
  States under the leadership of SRON Netherlands Institute for Space
  Research, Groningen, The Netherlands and with major contributions
  from Germany, France and the US. Consortium members are: Canada:
  CSA, U.Waterloo; France: CESR, LAB, LERMA, IRAM; Germany: KOSMA,
  MPIfR, MPS; Ireland, NUI Maynooth; Italy: ASI, IFSI-INAF,
  Osservatorio Astrofisico di Arcetri-INAF; Netherlands: SRON, TUD;
  Poland: CAMK, CBK; Spain: Observatorio Astron\'omico Nacional (IGN),
  Centro de Astrobiolog\'{\i}a (CSIC-INTA). Sweden: Chalmers
  University of Technology - MC2, RSS \& GARD; Onsala Space
  Observatory; Swedish National Space Board, Stockholm University -
  Stockholm Observatory; Switzerland: ETH Zurich, FHNW; USA: Caltech,
  JPL, NHSC.  We thank many funding agencies for financial support. 
  We would like to acknowledge S. Charnley, T. Jenness, R. Redman, R. Tilanus 
  and J. Wooterloot for their help in retrieving the para-D$_2$O data at JCMT. 
\end{acknowledgements}

\end{document}